\documentclass[a4paper,12pt]{article}
\usepackage{epsfig}
\usepackage{amssymb}
\usepackage{graphicx}

\usepackage{epsfig}
\usepackage{amssymb,amsmath}

\def\circa#1{\,\raise.3ex\hbox{$#1$\kern-.75em\lower1ex\hbox{$\sim$}}\,}

\newcommand{\GeV}{\,{\rm GeV}}
\newcommand{\TeV}{\,{\rm TeV}}

\makeatletter
\def\art{\@ifnextchar[{\eart}{\oart}}
\def\eart[#1]#2#3#4#5#6{{\rm #2}, {\em #3 \rm #4} {\rm (#6) #5 ({\em #1})}}
\def\hepart[#1]#2{{\rm #2, \em#1}}
\newcommand{\oart}[5]{{\rm #1}, {\em #2 \rm #3} {\rm (#5) #4}}

\newcommand{\beq}{\begin{equation}}
\newcommand{\eeq}{\end{equation}}
\newcommand{\bea}{\begin{eqnarray}}
\newcommand{\eea}{\end{eqnarray}}
\newcommand{\ba}{\begin{array}}
\newcommand{\ea}{\end{array}}
\newcommand{\bi}{\begin{itemize}}
\newcommand{\ei}{\end{itemize}}
\newcommand{\bn}{\begin{enumerate}}
\newcommand{\en}{\end{enumerate}}
\newcommand{\bc}{\begin{center}}
\newcommand{\ec}{\end{center}}

\newcommand{\eps}{\epsilon}
\newcommand{\al}{\alpha}
\newcommand{\be}{\beta}

\newcommand{\im}{{\rm Im}\,}
\newcommand{\re}{{\rm Re}\,}

\newcommand{\sla}[1]{\hspace{0.1cm} \diagup  \hspace{-0.3cm}{#1}}

\newcommand{\gsim}{\lower.7ex\hbox{$\;\stackrel{\textstyle>}{\sim}\;$}}
\newcommand{\lsim}{\lower.7ex\hbox{$\;\stackrel{\textstyle<}{\sim}\;$}}

\begin{document}
\tolerance=100000
\setcounter{page}{1}

\begin{flushright}
\vspace*{-2cm}
{UMD-PP-09-060\\ULB-TH/09-38}
\end{flushright}
\vspace{-0.1cm}
\begin{center}
{\Large\bf Reconciling leptogenesis with observable
$\mu \to e \gamma$ rates\\
\vspace{3mm}}
\end{center}
\vskip 0.5cm
\begin{center}
{\bf Steve~Blanchet$\,^{a}$,}\,
{\bf Thomas~Hambye$\,^{b}$,}\,
{\bf and}
{\bf Fran\c cois-Xavier~Josse-Michaux$\,^{b}$,}\,\footnote{sblanche@umd.edu, thambye@ulb.ac.be, fxjossemichaux@gmail.com}\\
\vskip .1cm
\vskip .7cm
$^a\,$ Department of Physics, University of Maryland, College Park, MD 20742, USA\\
\vspace{.2cm}
$^b\,$ Service de Physique Th\'eorique,
Universit\'e Libre de Bruxelles, 1050 Brussels, Belgium\\
\vskip .1cm
\end{center}
\vspace{0.2cm}

\begin{abstract}
We perform a detailed analysis of thermal leptogenesis in the
framework of seesaw models which approximately conserve lepton
number. These models are known to allow  for large Yukawa
couplings and a low seesaw scale in agreement with neutrino mass
constraints, and hence to lead to large lepton flavour violating rates that can be probed experimentally.
Although large Yukawa couplings lead to (inverse) decay
rates much larger than the Hubble expansion rate, we show that
the leptogenesis washout
induced is generically small if the mass splitting between the
right-handed neutrinos is small enough.
As a result, large lepton flavour violating rates
are compatible with successful leptogenesis. We emphasize that this
scenario does not require any particular flavour structure. A small
splitting is natural and radiatively stable in this context because
it is protected by the lepton number symmetry.

\vspace{-4mm}
\end{abstract}

\vspace{3mm}
\noindent
\section{Introduction}

The seesaw models
provide
an attractive and straightforward explanation for both the recently observed tiny neutrino masses and
the baryon asymmetry of the universe~\cite{Fukugita:1986hr}. However,
it is a very difficult task to probe these models experimentally,
especially the most  studied ``type-I" seesaw model
\cite{seesaw} based on the existence of gauge singlet right-handed
(RH) neutrinos.
In the generic situation, where \emph{lepton number} violation and \emph{lepton flavour} violation are generated at the same scale,
neutrino mass constraints require either an experimentally unreachable heavy scale for the mass of the RH neutrinos, or tiny Yukawa couplings (leading to suppressed RH neutrino production cross sections).
In both cases the rate of rare lepton processes, such as $\mu \rightarrow e \gamma$, is expected to be highly suppressed.

Without assuming any extra new physics beyond the right-handed neutrinos,\footnote{In the following we assume a non-supersymmetric setup. In supersymmetric seesaw models it is known that large flavour violating rates can be induced in agreement with leptogenesis, see
e.g.~\cite{ellrai,daviba}.}
there exists however a theoretically motivated class of seesaw models where
tiny neutrino masses do not imply that the lepton flavour violation is  suppressed: the seesaw models which approximately conserve the total lepton number
$L$~\cite{Wyler:1982dd}-\cite{Gavela:2009cd}.
These models are based on the fact that, even in the presence of
both RH neutrino masses and neutrino Yukawa couplings,
 it is possible to conserve
$L$, and thus to keep the
left-handed neutrinos massless. Hence the lepton flavour violating scale can be disconnected
from the lepton number violating one~(see \cite{Gavela:2009cd} for a detailed discussion).
As a result, the Yukawa couplings are not constrained to be small even if the RH neutrino mass scale is low. Large lepton-flavour-violating (LFV) --but $L$-conserving-- processes can be induced, with rates that could be observed
if the RH neutrino mass scale is sufficiently low, a few tens of TeV or less.
Neutrino masses can be explained subsequently by introducing small $L$-violating perturbations in the Yukawa coupling matrix and/or the RH neutrino mass matrix.

Successful baryogenesis via leptogenesis is known to be feasible at
the TeV scale ~\cite{Pilaftsis:1997,Pilaftsis:2003} through the
resonant leptogenesis mechanism
\cite{Flanz:1996fb,Covi:1996,Pilaftsis:1997}, which, through the
virtual $N$ propagator of the one-loop self-energy diagram
\cite{Liu:1993tg,Flanz:1994yx,Covi:1996wh}, requires a
quasi-degeneracy of at least two of the RH neutrinos. A nice feature
of the approximately $L$-conserving models is that they precisely
yield a quasi-degenerate spectrum, with a mass splitting
proportional
 to the small $L$-violating entries. Successful leptogenesis is achievable in this case, at least for Yukawa couplings suppressed
 enough~\cite{Asaka:2008}.

In this letter we investigate the following question: is successful leptogenesis compatible with observable rare
lepton processes? In Ref.~\cite{Pilaftsis:2005} it was shown to be feasible considering a particular flavour structure (suppressed
 $\tau$ Yukawa couplings) and with an extra $SO(3)$ symmetry between the 3 RH neutrinos. In the following we show that it is generically
  possible even with two RH neutrinos and without the need to assume any particular flavour structure.
This is due to the fact that the large Yukawa couplings which can lead to observable rare lepton processes are precisely the
ones which do not break lepton number,
therefore not necessarily causing any washout of the lepton asymmetry.

In Section 2 we introduce the model with approximate $L$ conservation which is going to be at the center of our study. In Section 3
we show explicitly how the washout is suppressed due to subtle interference effects in the $L$-conserving limit. In section 4 we compute the $C\!P$ asymmetry and, in Section 5, study the consequences
of our finding for the observability of rare lepton processes. In Section 6 we conclude
and emphasize the generality of our results.

\section{The model}

We consider for illustration the simple Type A model considered in~\cite{Gavela:2009cd} with only two RH neutrinos, $N_{1,2}$ (see also~\cite{Asaka:2008}).
With the charge assignment $L=1$, $L=-1$ to $N_1$ and $N_2$ respectively, the Lagrangian can be split into a $L$-conserving part and a
$L$-violating one:
\bea
\label{lagrangien}
\mathcal{L}_{L}&=&{\rm i} \overline{N}_{i}\sla{\partial}\,N_{i}-\left(Y_{\al}\,\overline{N}_{1}
\tilde{\phi}^{\dagger}\,\ell_{L\al}+\frac{1}{2} M\overline{N}_{1}N^c_{2}+h.c.\right)\,,\label{LC}\\
\mathcal{L}_{{\not}L}&=& -\left(Y^{\prime}_{\al}\,\overline{N}_{2}
\tilde{\phi}^{\dagger}\,\ell_{L\al}+\frac{1}{2}\mu_{1}\overline{N}_{1}N^c_{1}+\frac{1}{2}\mu_{2}{\rm
e}^{{\rm i}\alpha}\overline{N}_{2}N^c_{2}+h.c.\right)\,.\label{LV}
\eea with $\phi\equiv (\phi^+ \phi^0)^T$ the ordinary Higgs doublet.
Note that in this Lagrangian  $\mu_1$, $\mu_2$ and $M$ are real
parameters. The phases of $\mu_1$ and $M$ have been rotated away.
The Yukawa couplings $Y$ and the RH neutrino mass $M$ do not break
lepton number, i.e.~do not induce any $L$-violating dimension 5
operator for neutrino masses. But they do generate a dimension 6
$L$-conserving operator, and hence rare lepton processes, which can
be large. For example a $\mu \rightarrow e \gamma$ rate of order the
experimental upper bound Br$(\mu\rightarrow e \gamma)\simeq
10^{-11}$ can be obtained for $M \simeq$~1~TeV if the $Y$ couplings
are of order few $10^{-2}$. On the other hand, the couplings $Y'$,
as well as the diagonal mass entries $\mu_{1}$ and $\mu_2$, do break
lepton number and are therefore constrained to be small by the
neutrino masses: $Y'\ll Y$ and $\mu_{1},\,\mu_{2}\ll M$. This can be
seen by diagonalizing the neutral state mass matrix which results
from Eqs.~(\ref{LC})--(\ref{LV}) \bea \label{Massmatrix}
M_{\nu}=\left(
\begin{array}{ccc}
0 & Y_{\al}^{T}\,v& Y_{\al}^{' T}\,v\\
Y_{\al}\,v& \mu_{1} & M \\
Y_{\al}'\,v& M & \mu_{2}{\rm e}^{{\rm i}\alpha}
\end{array}\right)\,,
\eea  and which, at first order in $\mu_{1}$, $\mu_{2}$ and $Y'$,
gives the following light neutrino mass matrix (with $v= 174$~GeV):
\bea {m_\nu}= v^2 \left(Y'^T\frac{1}{M} Y+Y^T\frac{1}{M} Y'
\right)-v^{2}\,\left(Y^T\frac{1}{M}\mu_{2}{\rm e}^{{\rm
i}\alpha}\frac{1}{M} Y\right)\,.\label{mnu} \eea For $M\sim 1\TeV$ a
neutrino mass of order $0.1$~eV typically requires $Y'Y\sim
10^{-12}$ and/or $\mu_2 Y^2/M \sim 10^{-12}$. For $\mu_2=0$ this
model has also the nice and rather unique property to allow for the
full reconstruction of the flavour structure of the model from the
values of the light neutrino mass matrix
entries~\cite{Gavela:2009cd}. For $\mu_2 \neq 0$ it is not the case
but still the full flavour structure of the $L$-conserving rare
lepton processes can be reconstructed, leading to definite
predictions for the various rare lepton processes, up to an overall
normalization.

To consider leptogenesis in this framework, it is convenient to go
to the basis where the RH neutrino mass matrix is diagonal with real
and positive entries. The Lagrangian is then given by \bea
\label{lagrangien} \mathcal{L}&=&{\rm i}
\overline{N_{i}}\sla{\partial}\,N_{i}-\left(h_{i\al}\,\overline{N_{i}}
\tilde{\phi}^{\dagger}\,\ell_{L\al}+\frac{1}{2}
M_i\overline{N_{i}}N^c_{i}+h.c.\right),\quad(i=1,2;\,\al=e,\mu,\tau).\label{Lleptobasis}
\eea To first order in $\mu/M$ and $Y'$, the mass eigenvalues are
given by
\begin{equation}
M_{1,2}\simeq M \mp {1\over 2} \mu\,,
\end{equation}
where we defined $\mu_1+\mu_2 {\rm e}^{{\rm i}\alpha}\equiv \mu{\rm
e}^{{\rm i}\phi}$, so that
$\mu^2=\mu_1^2+\mu_2^2+2\mu_1\mu_2\cos\alpha$. As for the Yukawa
couplings, again to first order in $\mu/M$ and $Y'$, we find
\begin{eqnarray}
h_{1\alpha}&\simeq&{{\rm i}\over \sqrt{2}}{\rm e}^{-{\rm
i}(\phi-\lambda)/2}\left[\left(1+{\mu_2^2-\mu_1^2\over
4M\mu}\right){\rm e}^{{\rm i}\phi}
Y_{\alpha}-Y'_{\alpha}\right]\,,\label{h1}\\
h_{2\alpha}&\simeq&{1\over \sqrt{2}}{\rm e}^{-{\rm
i}(\phi+\lambda)/2}\left[\left(1-{\mu_2^2-\mu_1^2\over
4M\mu}\right){\rm e}^{{\rm i}\phi}
Y_{\alpha}+Y'_{\alpha}\right]\,,\label{h2}
\end{eqnarray}
where
\begin{equation}
\lambda=\sin\alpha {\mu_1\mu_2\over \mu M}.
\end{equation}

\section{Washout from $\Delta L=2$ scatterings}

The RH neutrinos being unstable particles, one cannot in principle define asymptotic free states for them. However, as shown in~\cite{Pilaftsis:1997,Buchmuller:1997}, their interaction properties (decay, inverse decay...) can be deduced from the $\Delta L=2$ scatterings mediated by on-shell heavy neutrinos $\ell_{\al} \tilde{\phi}^{\dagger}\leftrightarrow \bar{\ell_{\be}}\tilde{\phi}$. The  corresponding $\gamma_{\Delta L=2,\alpha}^{\rm tot}$ interaction rate enters in the Boltzmann equation for the lepton flavour $\alpha$ as
\bea
\label{Boltzmann}
s \,z\, H(z) \frac{d Y_{\ell_{\al}}}{dz}&=& \sum_{i=1}^{2} \eps_{i\al}\left(\frac{Y_{N_{i}}(z)}{Y_{N_{i}}^{eq}(z)}-1\right)\gamma_{N_{i}}^{D}-\sum_{i=1}^{2} \frac{Y_{\ell_{\al}}(z)}{Y_{\ell}^{eq}(z)}\,2\,\gamma_{\Delta L=2,i\al}^{\rm tot}\, ,
\eea
where $s$ is the entropy density, $z\equiv M_{1}/T$, $H(z)$ is the Hubble expansion rate,  $Y_X\equiv n_X/s$ is the number density of particle $X$ normalized to
the entropy density and  $\epsilon_{i\alpha}$ is the $C\!P$ asymmetry from the RH neutrino $N_i$
for the flavour $\alpha$.
Note that we consider a Boltzmann equation for each flavour, even though our results will not require to assume any particular flavour pattern.

One can separate the $\Delta L=2$ scattering into an on-shell and off-shell contribution,
$\gamma_{\Delta L=2,\alpha}^{\rm tot}=\gamma_{\Delta L=2,\alpha}^{\rm on}+\gamma_{\Delta L=2,\alpha}^{\rm off}$. 
As shown in the appendix of~\cite{Giudice:2003}, in usual
leptogenesis scenarios, the on-shell part of the $\Delta L=2$
scattering, which is the dominant contribution, yields the inverse
decays of leptons into heavy neutrinos, \bea \gamma_{\Delta
L=2,i\al}^{\rm on} = \frac{\gamma^D_{N_i,\al}}{4}\equiv
\frac{1}{4}\, n_{N_i}^{\rm
eq}\,\frac{\mathcal{K}_{1}}{\mathcal{K}_2}
\,\Gamma_{N_i,\al}=\frac{4}{z} \frac{M_i^4}{128\pi^3}|h_{i\al}|^2
\mathcal{K}_{1}(z\, M_i/M_1) \label{gammaD} \eea where
$\mathcal{K}_{1,2}(z)$ are the modified Bessel functions. Note that
$\sum_{\al}\gamma^D_{N_i,\al}=\gamma^D_{N_i}$. Inverse decays
contribute to a depletion of the lepton asymmetry, parametrized by
the washout parameter $K_{i\al}\equiv\Gamma_{N_i,\al}/H(T=M_{N_i})$,
where the $N_i$ decay width is $\Gamma_{N_i,\al}= \vert
h_{i\al}\vert^{2}\,M_{i}/8\,\pi$. With quasi-degenerate  RH
neutrinos the relevant washout parameter is given by the sum
$K_\alpha \equiv K_{1\alpha}+K_{2\alpha}= \vert
Y_{\al}\vert^{2}\,M/8\,\pi/H(M) +{\cal O}(Y'^2,\mu^2/M^2)$.

There is an apparent contradiction here. On the one hand, it is
clear that in the $L$-conserving limit $\Delta L=2$ scatterings must
vanish. On the other hand, the on-shell contribution quoted above
clearly does not vanish, since the decay width is dominated by the
$L$-conserving $Y$ couplings, $\vert h \vert^{2} \simeq |Y|^2$. For
example, if the $Y$ couplings are large, say $10^{-2}$, and if the
RH neutrino mass $M$ is as low as 1 TeV, the washout parameter
$K_{\al}$ is huge, about $10^9$, which gives a baryon asymmetry way
too small.
However, this huge washout is independent of the $L$-violating parameters, i.e. it
would be present even in the absence of $L$-violation, which is clearly erroneous.

In the following we show that this apparent contradiction disappears when one takes into account, on top of the inverse decay terms, the usually neglected interference terms involving different $N_i$'s.
To this end we start from the $\Delta L=2$ scattering reaction
density as given from the reduced cross section
$\hat{\sigma}_{\Delta L=2,\alpha}$: \bea \gamma_{\Delta
L=2,\al}^{\rm tot}(z)=
\frac{1}{z}\frac{M_{1}^{4}}{64\,\pi^{4}}\,\int_{x_{\rm
thr}}^{\infty}dx\,\sqrt{x}\,\hat{\sigma}_{\Delta
L=2,\al}(x)\,\mathcal{K}_{1}(z\,\sqrt{x})\,, \label{gammadelL2} \eea
where $x\equiv s/M_{N_1}^2$, and with the scattering threshold given
by $x_{\rm thr}=(m_h+m_{\ell})^2/M^2\simeq 0$.
The $s$-channel reduced cross section is given by\footnote{Here for
simplicity we only consider the on-shell part, but it can be easily
shown that the total $\Delta L=2$ rate vanishes in the
$L$-conserving limit, and therefore so does the off-shell part.}
\bea
\hat{\sigma}_{\Delta L=2,\al}= \vert
h_{1\al}\vert^{2}A_{11}\sum_{\be}\vert h_{1\be }\vert^{2} +\vert
h_{2\al }\vert^{2}A_{22}\sum_{\be}\vert h_{2\be
}\vert^{2}+\,2\,\re\left[ \left( h_{1\al }h_{2\al
}^{\star}\right)A_{12}\sum_{\be}\left( h_{1\be
}h_{2\be}^{\star}\right)\right],
\eea where the $h_{\al i}$'s are
the neutrino Yukawa couplings of Eqs.~(\ref{h1})--(\ref{h2}). In the
above expression \bea A_{ij}=\frac{x\,\sqrt{a_{i}a_{j} } }{4\pi
P_{i}P_{j}^{\star} }\,, \eea with $P_{i}^{-1}=
\left(x-a_{i}+i\sqrt{a_{i} c_{i}}\right)^{-1}$ the dressed inverse
propagators, whose poles are located at
$\hat{a}_{i}=a_{i}-i\sqrt{a_{i}c_{i}}$, with
$a_{i}=(M_{i}/M_{1})^{2}$ and $c_{i}=(\Gamma_{i}/M_{1})^{2}$. In the
following, whenever the small $L$-violating $c_2-c_1$ difference is
irrelevant we will take $c_1=c_2=c$.

Using a narrow width approximation,\footnote{Or equivalently, up to negligible terms, taking the residue of the integral in Eq.~(\ref{gammadelL2}) at the physical poles $\hat{a}_{1,2}$.} the $\Delta L=2$ washout term is given by
\bea
\gamma_{\Delta L =2,\al}^{\rm on}&=&
\frac{1}{z}\frac{M_{1}^{4}}{64\,\pi^{4}}\frac{1}{4\pi}\, \mathcal{K}_{1}
\left(z\right) |Y_{\alpha}|^2\,\sum_{\beta}|Y_{\beta}|^2 \\
&\times & \left[{a_1^2+a_2^2\over \sqrt{c}}-{2\sqrt{a_1a_2}(\sqrt{a_1c_1}+\sqrt{a_2c_2})\over (a_2-a_1)^2+(\sqrt{a_1c_1}+\sqrt{a_2c_2})^2}\left(a_1^{3/2}+a_2^{3/2}\right)
+{\cal O}(Y'^{2},\mu^{2})\right]\,.\nonumber
\label{gammadeltaA}
\eea
The first term in the parenthesis of Eq.~(\ref{gammadeltaA}) arises from the pure $N_{1}$ and $N_2$ contributions, and yields inverse decay terms as
in Eq.~(\ref{gammaD}). The second term is an interference effect, which is usually neglected in standard seesaw models. However, in the models we consider, this cannot be done, since both contributions are of the same order.

Factoring out the $L$-conserving contribution to the
$\gamma^{D}_{N_{1,2,\al}}$ inverse decay terms, which up to ${\cal
O}(Y'^2,\mu^2/M^2)$ terms is given by $\gamma^{D}_{N,\al}\equiv
(\gamma^{D}_{N_1,\al}+\gamma^{D}_{N_2,\al})/2$, one finally obtains
for the total washout \bea \gamma_{\Delta L =2,\al}^{\rm
on}=\frac{\gamma^{D}_{N,\al}}{4} \cdot 2
\left(1+2\delta\sqrt{c}-{1+3\delta\sqrt{c}\over
1+\delta\sqrt{c}+\delta^2} +{\cal O}(Y'^{2},\mu^{2})\right) =
\frac{\gamma^{D}_{N,\al}}{4}\cdot
\frac{2\,\delta^{2}}{1+\sqrt{c}\delta+\delta^{2}}\, ,
\label{interferencefinalN1} \eea where the mass splitting enters in
the parameter $\delta \equiv (M_{2}-M_{1})/\Gamma_1=\mu/(M\sqrt{c})
\simeq (a_{2}-a_{1})/2\sqrt{c}$.  There is an obvious cancellation
in the parenthesis between the first two terms (pure $N_1$ and $N_2$
contributions) and the third one ($N_1$-$N_2$ interference), leaving
a washout which vanishes in the $L$-conserving limit, $\delta
\rightarrow 0$. Therefore, no matter how huge the ``naive'' inverse
decay term in Eq.~(\ref{gammaD}) is, there is no washout in the
$L$-conserving limit, as expected. Conversely, for a large mass
splitting $\delta \gg 1$, the interference term becomes negligible,
and one recovers the usual washout from inverse decays. This shows
that the interference term which is rightly neglected in typical
seesaw models (see e.g. \cite{Giudice:2003,Buchmuller:2004}) can be
essential in the context of approximate $L$-conserving frameworks.

Practically, since the washout vanishes in the $L$-conserving limit,
it remains small as long as the $L$-violating perturbations, $Y'$,
$\mu_{1,2}$, remain small enough: it is suppressed as long as
$\delta \ll 1$. The relevant washout parameter including the
interference term turns out to be \beq\label{Keff} K_{\al}^{\rm eff}
\equiv K_{\alpha} \cdot
\frac{\delta^{2}}{1+\sqrt{c}\delta+\delta^{2}} \,\stackrel{\delta\ll
1}{\simeq}\,
 K_{\alpha} \cdot \delta^2\,.
\eeq
For the numerical example above with $M\sim 1$~TeV and $K_{\al}\sim 10^9$, if $\delta \lesssim 10^{-3}$
one ends up with an effective washout $K^{\rm eff}_{\al}\simeq K_{\alpha} \cdot \delta^{2}\lesssim 10^{3}$,
which is suppressed enough to keep a substantial lepton asymmetry.

To sum up,
in the class of models we are considering, the washout of the asymmetry can be suppressed by many orders of magnitude, granted that the
degeneracy parameter $\delta$ is small enough.
A splitting $\delta \ll 1$ is technically natural since it is controlled by $L$-violating parameters. As a matter of fact,
the radiative and thermal corrections to $\delta$ are proportional to $YY'$ and not to $Y^2$ as in typical seesaw scenarios.
It is interesting to notice that, contrary to usual seesaw models with two RH neutrinos where the total washout is lower-bounded by the
solar scale
$K_{1}+K_{2} \gtrsim K_{\rm sol}\sim 9$, the effective washout in Eq.~(\ref{Keff}) is lower-bounded only by terms quadratic in the small
$L$-violating parameters $Y'$ and $\mu/M$. Therefore, $K_{\alpha}^{\rm eff} \ll K_{\rm sol}$ can be
achieved in this case.

\section{CP asymmetry}

The $C\!P$ asymmetry generated during the decays of $N_{i}$ into the
lepton flavour $\al$ is given by~\cite{Covi:1996} \bea
\eps_{i\al}=\frac{1}{8\pi}\sum_{j\neq i}\left\lbrace
{\im{\left[h_{i\al}^{*} h_{j\al} \left(\sum_{\gamma} h_{i\gamma}^{*}
h_{j\gamma }\right)\right]} \over \sum_{\be} \vert h_{i\be}\vert^{2}
} f_{v}^{i,j}+{\im{\left[h_{i\al }^{*} h_{j\al}\left(\sum_{\gamma}
h_{i\gamma } h_{j\gamma}^{*} \right)\right]} \over \sum_{\be} \vert
h_{i\be}\vert^{2} } f_{c}^{i,j}\right\rbrace , \eea where $f_{v}$ is
the usual $L$-violating self-energy and vertex loop factor and
$f_{c}$ is a $L$-conserving self-energy loop factor. In the limit we
are interested in, namely $M_{2}\simeq M_{1}$, only the self-energy
correction is relevant, and we get
$f_{v}^{2,1}\simeq-f_{v}^{1,2}\simeq f_{c}^{2,1}\simeq
-f_{c}^{1,2}\equiv f_{\rm self}$ with
~\cite{Buchmuller:1997,Anisimov:2005} \bea f_{\rm
self}=\frac{a_{2}-a_{1}}{\left(a_{2}-a_{1}\right)^{2}+\left(\sqrt{a_{2}c_{2}}-\sqrt{a_{1}c_{1}}\right)^{2}}\,.
\label{regulator} \eea It is instructive to express the $C\!P$
asymmetry in terms of the parameters of the original Lagrangian,
Eqs.~(\ref{LC})--(\ref{LV}):
 \bea
 \eps_{1\al} =\eps_{2\al} \simeq -
{|Y_{\alpha}|^2\over 4\pi}\left(\sin\alpha {\mu_1\mu_2\over 2\mu M}+
{\sum_{\beta}{\rm Im}(Y_{\beta}Y_{\beta}'^{*} {\rm e}^{{\rm
i}\phi})\over \sum_{\beta'}|Y_{\beta'}|^2}\right)\,f_{\rm self}\, .
\label{CPasymm} \eea The above expression shows explicitly that the
$C\!P$ asymmetry in each flavour crucially depends on the
$L$-violating parameters, and therefore vanishes when $L$ is
conserved.\footnote{In models with more fermion singlets, a term in
the self-energy diagram may remain in the limit of $L$ conservation,
see e.g.~$B-L$ conserving models of
leptogenesis~\cite{GonzalezGarcia:2009qd} or models with an extra
$N_3$ \cite{Pilaftsis:2005,Antusch:2009gn}. Unless an extra symmetry
beyond $L$ conservation is assumed to have a quasi-degeneracy
between at least 3 fermion singlets, such a term turns out to be
generically irrelevant for the generation of an asymmetry at the TeV
scale due to lepton flavour equilibration \cite{Antusch:2009gn}.}

At the singularity $M_{1}=M_{2}$, the self-energy correction above,
Eq.~(\ref{regulator}), has a regulator given by
$\sqrt{a_{2}c_{2}}-\sqrt{a_{1}c_{1}}$. In usual seesaw scenarios,
one typically has $|\sqrt{a_{2}c_{2}}-\sqrt{a_{1}c_{1}}|~\simeq~{\rm
max_i}(\sqrt{a_{i}c_{i}})$ because the decay widths
$\Gamma_{N_{1,2}}$ are very different from each other. However, in
our case, while the $\sqrt{a_{i}c_{i}}$ do not vanish in the
$L$-conserving limit, $\sqrt{a_{2}c_{2}}-\sqrt{a_{1}c_{1}}$ does. As
a result $|\sqrt{a_{2}c_{2}}-\sqrt{a_{1}c_{1}}| \ll {\rm
min_i}(\sqrt{a_{i}c_{i}})$. Therefore the resonance, which occurs
for $a_2-a_1~=~\sqrt{a_{2}c_{2}}-\sqrt{a_{1}c_{1}}$, \textit{i.e.}
for $\delta=\delta_{\rm res}  \simeq~2\,{\rm Re}(Y{Y'}^* {\rm
e}^{{\rm i}\phi})/\vert Y\vert ^2 -(\mu_2^2-\mu_1^2)/2M\mu \ll 1$,
is obtained for a mass splitting much smaller than the decay width.
Around the resonance, where the $C\!P$ asymmetry reaches its maximum
value $\eps\sim 1$, the $C\!P$ asymmetry is no longer suppressed by
the small $L$-violating parameters. In practice in the following we
will never need to use such a small mass splitting, since in this
case, unless the $C\!P$ phases are highly suppressed, the baryon
asymmetry produced by leptogenesis would be far bigger than the
observed value. Therefore, we will always stick to the natural range
$\vert \sqrt{a_{2}c_{2}}-\sqrt{a_{1}c_{1}}\vert \ll a_{2}-a_{1} \ll
{\rm max_{i}}(\sqrt{a_{i}c_{i}})$, or equivalently $\delta_{\rm res}
\ll \delta \ll 1$. In this regime, the self-energy contribution is
simply given by \beq\label{selfCP} f_{\rm self}\simeq {1 \over
2\delta \sqrt{c}}. \eeq It is remarkable that the conditions to get
a large enough $C\!P$ asymmetry and a suppressed washout are the
same: $\delta \ll 1$.

Note that, using a different method, a regulator $\propto
\sqrt{a_{i}c_{i}}$ was obtained in
\cite{Pilaftsis:2003,Pilaftsis:2005}, giving a resonance peak for
$\delta\sim 1$. In the following we will stick to the regulator
found in \cite{Buchmuller:1997,Anisimov:2005},
Eq.~(\ref{regulator}), for two reasons. First, in
\cite{Anisimov:2005}, the self-energy contribution was calculated
using two quantum field theory methods, and both converge to the
regulator in Eq.~(\ref{regulator}). Second, the very different
oscillation formalism typically applying to the
$K^{0}$--$\overline{K}^{0}$ and $B^0$--$\overline{B}^0$ systems was
used in~\cite{Covi:1996} to compute the leptogenesis self-energy
diagram in the limit of small mass splitting and the same result was
obtained. Note also that the results of \cite{Anisimov:2005},
Eq.~(\ref{regulator}), are obtained in the limit where the
off-diagonal corrections to the propagator in the $N_{1}$--$N_{2}$
system are small compared to the diagonal ones. More precisely, in
terms of the parameters of the model, the condition of
Ref.~\cite{Anisimov:2005} reads $[(h^{\dagger}h)_{12}
+(h^{\dagger}h)_{21}]/(32\pi^2)\ll (M_2-M_1)/M_1$. In our framework,
since the off-diagonal corrections are proportional to the
$L$-violating parameters, i.e. $(h^{\dagger}h)_{12}
+(h^{\dagger}h)_{21}\propto {\cal O}(Y'Y,\mu/M)$, this condition is
clearly fulfilled.

\section{Leptogenesis vs.~rare lepton processes}

It is well known that for RH neutrino masses of order the weak scale and Yukawa couplings of order $ {\rm few} 10^{-2}$,
rates for rare lepton processes such as $\mu \to e\gamma$
close to the present experimental bound, ${\rm Br}(\mu\rightarrow e \gamma)\lesssim 1.2\times 10^{-11}$~\cite{PDG}, can be obtained \cite{Cheng:1980tp},\cite{Antusch:2006vwa}.
From the suppression of the washout in Eq.~(\ref{Keff}) and the form of the $C\!P$ asymmetry shown in Eqs.~(\ref{CPasymm}) and (\ref{selfCP}),
it is easy to convince oneself that successful leptogenesis is possible at the same time.
Taking for example $M\sim 1$~TeV, $Y\sim \, \hbox{few}\,10^{-2}$, $Y'\sim Y\mu/M\sim 10^{-10}$ (which may lead easily to the correct neutrino masses pattern), one gets $\delta \sim 10^{-5}$
and thus a $C\!P$ asymmetry $\epsilon\sim {\rm few\,} \,10^{-6}$. In this case,
even though $K_{\alpha}\sim 10^{11}$, the effective washout in
Eq.~(\ref{Keff}) is reduced by a factor $1/\delta^2\sim 10^{10}$, so that $K^{\rm eff}\sim 10^2$. Such values of the washout and of the $C\!P$
asymmetry are in the right ballpark to explain the observed baryon asymmetry of the Universe.

To study this result more quantitatively it is convenient to use the
Casas-Ibarra parametrization of the Yukawa
couplings~\cite{Casas:2001}, which, in the 2 heavy neutrino case and
for a normal hierarchy in the light sector, reads \bea\label{casas}
h_{1\al}&=&\frac{\sqrt{M_{1} } }{v}\left( \sqrt{m_{2}}\,\cos{z}\,U_{\al 2}^{\star}+\,\sqrt{m_{3}}\,\sin{z}\,U_{\al 3}^{\star}\right)\,, \\
h_{2\al}&=&\frac{\sqrt{M_{2} } }{v}\left(
-\sqrt{m_{2}}\,\sin{z}\,U_{\al
2}^{\star}+\,\sqrt{m_{3}}\,\cos{z}\,U_{\al 3}^{\star}\right)\,. \eea
The inverted hierarchy is obtained simply by exchanging the indices
$2\to 1$ and $3\to 2$. In this parametrization, $z=z_{a}+{\rm i}
z_{b}$ is a complex angle, and $U$ is the PMNS matrix. We will adopt
the parametrization

\begin{equation}\label{Umatrix}
U=\left( \begin{array}{ccc}
c_{12}\,c_{13} & s_{12}\,c_{13} & s_{13}\,{\rm e}^{-{\rm i}\,\delta_{{\rm CP}}} \\
-s_{12}\,c_{23}-c_{12}\,s_{23}\,s_{13}\,{\rm e}^{{\rm i}\,\delta_{{\rm CP}}} &
c_{12}\,c_{23}-s_{12}\,s_{23}\,s_{13}\,{\rm e}^{{\rm i}\,\delta_{{\rm CP}}} & s_{23}\,c_{13} \\
s_{12}\,s_{23}-c_{12}\,c_{23}\,s_{13}\,{\rm e}^{{\rm i}\,\delta_{{\rm CP}}}
& -c_{12}\,s_{23}-s_{12}\,c_{23}\,s_{13}\,{\rm e}^{{\rm i}\,\delta_{{\rm CP}}}  &
c_{23}\,c_{13}
\end{array}\right)
\cdot {\rm diag(1, {\rm e}^{{\rm -i}\,{\gamma / 2}}, 1)} ,
\end{equation}

where $s_{ij}\equiv \sin\theta_{ij}$, $c_{ij}\equiv\cos\theta_{ij}$.
Neglecting experimental errors, we will use
$\sin^2(2\theta_{12})~=~0.87$,
$\sin^2(2\theta_{23})~=~0.92$~\cite{PDG},
$\sin^2(\theta_{13})~<~0.03\,\,(1\sigma)$~\cite{Fogli:2009ce},
whereas~for~the neutrino mass-squared differences we will take
$\Delta m_{\rm sol}^2 \simeq 7.59 \times 10^{-5}$~eV$^{2}$ and
$\Delta m_{\rm atm}^2 \simeq 2.43 \times
10^{-3}$~eV$^{2}$~\cite{PDG}. Recall that with only two RH neutrinos
the lightest active neutrino is massless. Therefore, if the neutrino
mass hierarchy is normal, one has $m_1=0$, $m_2=\sqrt{\Delta m_{\rm
sol}^2}$ and $m_3=\sqrt{\Delta m_{\rm sol}^2 + \Delta m_{\rm
atm}^2}$, whereas if it is inverted, $m_3=0$, $m_2=\sqrt{\Delta
m_{\rm atm}^2}$ and $m_1=\sqrt{\Delta m_{\rm atm}^2 - \Delta m_{\rm
sol}^2}$.

In order to use the parametrization of Eq.~(\ref{casas}) in a regime
corresponding to the model we consider, Eqs.~(\ref{LC})--(\ref{LV}),
let us invert Eq.~(\ref{h1}) to write the couplings $Y$ and $Y'$ as
\bea Y_{\al}&=&{\rm e}^{-{\rm i}\phi/2}\sqrt{\frac{M}{2}}\frac{{\rm
e}^{z_b}}{v}\,{\rm e}^{-{\rm i}\,z_a}\left(-{\rm i}
\sqrt{m_{2}}\,U_{\al 2}^{*}+\sqrt{m_{3}}\,U_{\al 3}^{*}\right)\, ,\\
Y_{\al}'& =& {\rm e}^{{\rm i}\phi/2}\sqrt{\frac{M}{2}}\frac{{\rm
e}^{-z_b}}{v} {\rm e}^{{\rm i}\,z_a}\left({\rm
i}\sqrt{m_{2}}\,U_{\al 2}^{*}+\sqrt{m_{3}}\,U_{\al 3}^{*}\right)+
{Y_{\alpha}{\rm e}^{{\rm i}\phi}\over 2}\left({\mu_2^2-\mu_1^2\over
2 M\mu}+{\rm i}\lambda\right)\,. \eea It is clear that the
parametrization above yields $Y'\ll Y$ if ${\rm e}^{-z_{b}}\ll 1$,
the $\mu/M$ term being anyway suppressed by the approximate lepton
number symmetry. In this limit, the $C\!P$ asymmetry,
Eq.~(\ref{CPasymm}), takes the simple form: \bea
\eps_{\al}=4\,P_{\al}\,\frac{m_{3}-m_{2}}{m_3+m_2}\,\sin{(2z_{a})}\,\frac{{\rm
e}^{-2\,z_{b}}}{\delta} \label{EcpFin} \eea where \bea
P_{\al}=\frac{m_{3}\vert U_{\al 3}\vert^2+2 \sqrt{m_{3}m_{2}}\,{\rm
Im}(U_{\al 2}^{*}U_{\al 3})+m_{2}\vert U_{\al
2}\vert^{2}}{m_{3}+m_{2}}\, \eea is the branching ratio of the decay
of $N_{i}$ into the flavour $\al$, with $\sum_{\al} P_{\al}=1$.
$C\!P$ violation in Eq.~(\ref{EcpFin}) is controlled by the
``leptogenesis phase'' $z_a$ and is maximal for $z_{a}=\pi/4$. It is
interesting to notice that the $C\!P$ asymmetry in the case of
inverted hierarchy is suppressed compared to normal case by a factor
$\sqrt{\Delta m_{\rm sol}^2/\Delta m_{\rm atm}^2}$. As for the
$K_{\al}^{\rm eff}$ washout parameter,  Eq.~(\ref{Keff}), for
$\delta \ll 1$ it is simply given by \bea K_{\al}^{\rm eff}\simeq
\frac{M}{H(M)}\frac{\vert
Y_{\al}\vert^{2}}{8\pi}\,\delta^{2}=P_{\al}\,
\frac{m_{3(2)}+m_{2(1)}}{2 m_{\star}}\,\,{\rm
e}^{2\,z_{b}}\delta^{2}\,\simeq \chi_1\,P_{\al}\,{\rm
e}^{2\,z_{b}}\delta^{2}\,, \eea where $m_{\star}\simeq 1.08\times
10^{-3}$~eV and $\chi_1\simeq 27\,(45)$ for normal (inverted)
hierarchy. Had we neglected the interference term, we would have
obtained a huge washout, $K_{\al} \simeq \chi_{1}\,\,e^{2\,z_{b}}$.
Though the suppression of the washout $\propto \delta^2$ is crucial,
it turns out that we will always remain in the regime of ``strong''
washout, defined as $K^{\rm eff}_{\alpha}\gtrsim 3$. In this regime,
where the dependence on the initial conditions disappears, and where
subtle effects like thermal corrections are
irrelevant~\cite{Giudice:2003}, we obtain finally for the baryon
asymmetry \bea \label{Yb}
Y_{B}=\frac{12}{37}\sum_{\al}0.5\frac{\eps_{\al}}{(K_{\al}^{\rm
eff})^{1.16}}\,Y_{N}^{\rm eq}(T\gg M)\simeq
\chi_2\,\left(\sum_{\al}P_{\al}^{-0.16}\right)\,\frac{e^{-4.32\,z_{b}}}{\delta^{3.32}}\sin{(2z_{a})}\,,
\eea where $\chi_2\simeq 8.7\times 10^{-5}$ ($5.4\times 10^{-7}$)
for normal (inverted) hierarchy. There is a mild dependence on the
parameters of the PMNS matrix in $\sum_{\al}P_{\al}^{-0.16}$, which,
barring large cancellations, varies between 3.7 and 4.8 for normal
hierarchy and 3.6 and 4 for inverted. In the following we will
constrain the baryon asymmetry to lie within the $3\sigma$ range of
WMAP5, i.e. $8.1\times 10^{-11}<Y_{B}<9.5\times
10^{-11}$~\cite{WMAP5}.

Let us now turn to the low-energy observables. The branching ratio for the rare lepton process $\mu \to e\gamma$ is given by~\cite{Cheng:1980tp}:
\bea
{\rm Br}(\mu\rightarrow e \gamma)&\simeq &\frac{3\alpha}{32\pi} \left(2\,v^{2}\right)^{2}\left|
\frac{ h_{1e}h_{1\mu}^{\star}}{M_{1}^{2}} +\frac{h_{2e}h_{2\mu}^{\star} }{M_{2}^{2}}\right|^{2}\, ,\nonumber\\
&\simeq&
\frac{3\alpha}{32\,\pi}\left(\frac{m_{3}+m_{2}}{M}\right)^{2}\,\vert
g_{e\mu}\vert^{2}\,{\rm e}^{4z_{b}}\,, \eea where $\alpha$ is the
fine structure constant, and \beq g_{e\mu}\equiv \frac{m_{2}\,
U_{e2}^{\star}U_{\mu 2}+m_3\, U_{e3}^{\star}U_{\mu 3}+2\,{\rm
i}\sqrt{m_2 m_3}(U_{e3}^{\star}U_{\mu 2}-U_{e 2}^{\star} U_{\mu
3})}{m_3+m_2}. \eeq As before, the replacement $3\to 2$ and $2\to 1$
should be made for the case of inverted hierarchy. The dependence on
the parameters of the PMNS matrix is clearly larger for the
branching ratio than for the baryon asymmetry. Here, $|g_{e\mu}|^2$
varies from 0.01 to 0.2 for the normal case, and from 0.06 to 0.7 in
the inverted one, also granted that no large cancellations occur.

Using Eq.~(\ref{Yb}) successful leptogenesis implies a relation between $z_{b}$ and the degeneracy parameter $\delta$, which we use to rewrite
the branching ratio as
\bea
{\rm Br}(\mu\rightarrow e \gamma)&\simeq &\frac{3\,\alpha}{32\,\pi}\left(\frac{m_{3}+m_{2}}{M}\right)^{2}\,\vert g_{e\mu}\vert^{2}
\left(\frac{\chi_{2}\,\left(\sum_{\al}P_{\al}^{-0.16}\right)\sin{(2z_{a})}}{Y_{B}^{\rm obs}\,\delta^{3.32}} \right)^{4/4.32}\,\nonumber \\
&\simeq& \chi_{3}\times
10^{-12}\left(\frac{250\GeV}{M}\right)^{2}\times\left(\frac{10^{-4}}{\delta}\right)^{3.1}\times\sin{(2z_{a})}^{0.93}\,,
\label{muegamestim} \eea with $\chi_{3}$ taking values between
$4.1\times 10^{-1}$ and $7.6$ for the normal case and between
$4.2\times 10^{-2}$ and $2.9\times 10^{-1}$ in the inverted one.
From the above equations it is clear that the smaller the degeneracy
parameter $\delta$ is, the bigger the Yukawa couplings can be
without spoiling the success of leptogenesis, hence the larger the
LFV rate $\mu\rightarrow e \gamma$ is. Eq.~(\ref{muegamestim}) shows
that for $M\sim 250\GeV$--$1\TeV$, observable $\mu\rightarrow e
\gamma$ rates typically require $\delta \sim 10^{-3}$--$10^{-5}$.
This corresponds to a mass splitting $(M_{2}-M_{1})/M_{1}\sim
10^{-8}$--$10^{-10}$. Such a level of degeneracy is comparable to
the one needed in usual resonant leptogenesis scenarios.

To check that our analytical result of Eq.~(\ref{muegamestim}) is
correct, it is necessary to integrate numerically the set of
Boltzmann equations, as given in Eq.~(\ref{Boltzmann}). Fig.~1 shows
our numerical results for the simple case $\mu_1=0$, $M=250\GeV$,
$\delta_{\rm CP}=\gamma=0$ and $\theta_{13}=10^{\circ}$, taking for
the leptogenesis phase $z_a$ its maximum value $\pi/4$. In this
simple case, as $\mu_1=0$, the phase $\alpha$ is unphysical and only
the second term of Eq.~(\ref{CPasymm}) contributes. The parameter
$\mu_2$ is directly responsible for the mass splitting between the
RH neutrinos, while $z_b$ sets the scale of the Yukawa couplings.
Fig.~1 displays, as a function of these 2 parameters, the
iso-contours of the LFV rate compatible with successful leptogenesis
for both normal and inverted neutrino mass hierarchies. A sphaleron
freeze-out cut-off has been applied at the temperature $T\sim
130\GeV$ taking the Higgs boson mass
$m_h=120$~GeV~\cite{Burnier:2005}. We find a good agreement
with the estimate of Eq.~(\ref{muegamestim}), up to the sphaleron
freezeout which has not been taken into account in this equation.
Also given in Fig.~1 are the LFV contour plot as a function of the
parameters $K^{\rm eff}$ and $\delta$. We observe that the branching
ratio reached can saturate or even exceed the present experimental
limit, ${\rm Br}(\mu\rightarrow e \gamma)\lesssim 1.2\times
10^{-11}$, depicted as a green line in Fig.~1.
\begin{figure}[t!]
\label{Graphe1}
\hspace{-0.5cm}
\includegraphics[width=17cm]{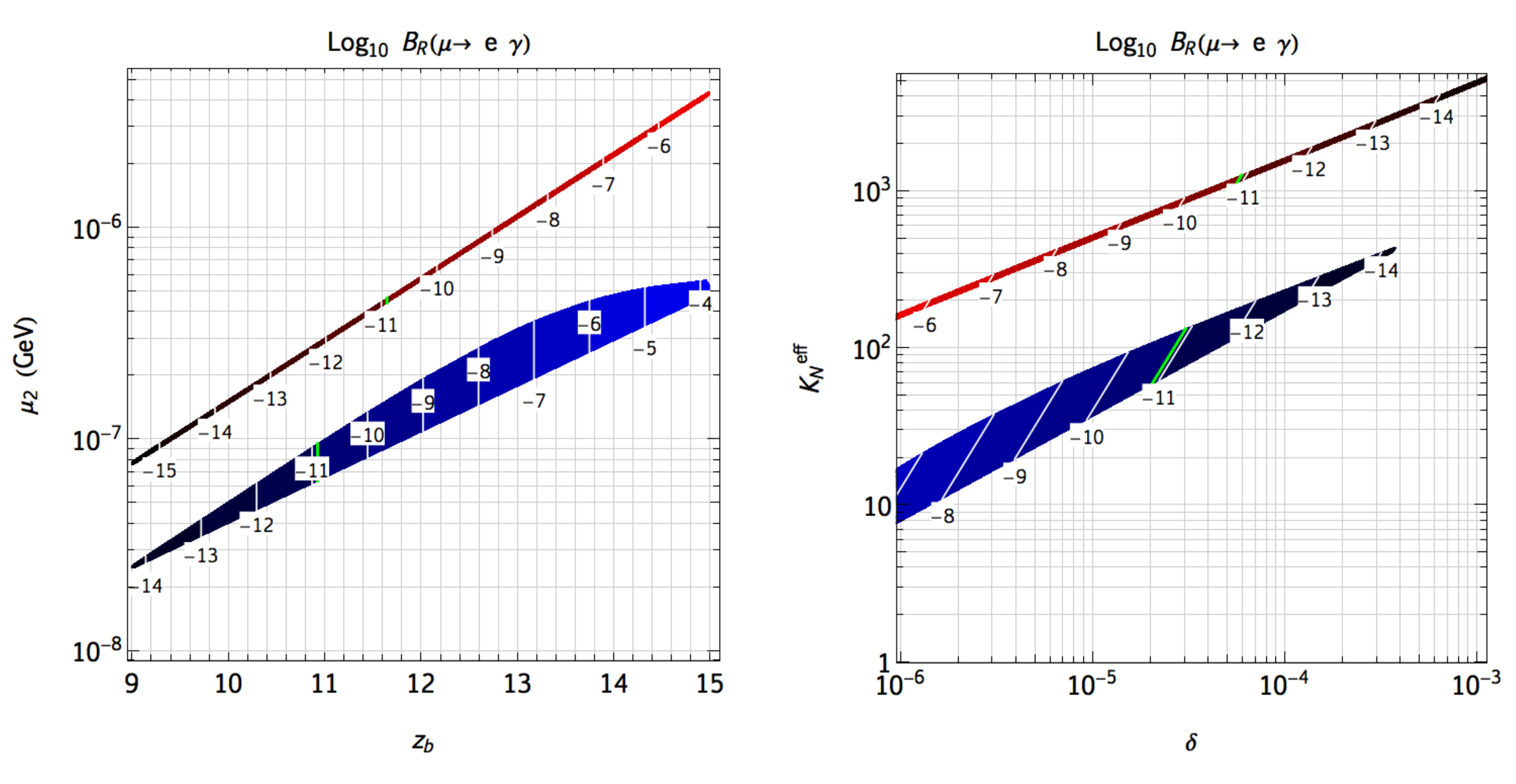}
\caption{Contours of ${\rm Br}(\mu\rightarrow e \gamma)$ compatible with successful leptogenesis plotted against $\mu_{2}$ and $z_{b}$ (left panel), and
$K^{\rm eff}$ and $\delta$ (right panel). The upper (lower) values correspond to the normal (inverted) hierarchical neutrino spectrum. The green
line shows the present experimental limit ${\rm Br}(\mu\rightarrow e \gamma)\lesssim 1.2\times 10^{-11}$.}
\end{figure}

The $\mu \rightarrow e \gamma$ decay is not the only flavour
violating process whose current experimental upper bound could be
saturated in agreement with leptogenesis constraints. This is also
possible for the $\tau$ radiative decays, which experimentally are
constrained to the bounds ${\rm Br}(\tau \rightarrow \mu \gamma)<
4.5\cdot 10^{-8}$ and ${\rm Br}(\tau \rightarrow e \gamma)< 1.1
\cdot 10^{-7}$ \cite{PDG}. It turns out that the results of Fig.~1
can be applied
to these rates for a degeneracy parameter $\delta$ about one
order of magnitude smaller than for the
$\mu \rightarrow e \gamma$ rate\footnote{The Yukawa couplings
in the $\mu$-$e$ channel should be somewhat suppressed with respect to the
ones of the other channel(s) for the more stringent $\mu\rightarrow
e \gamma$ constraint to be satisfied.}. As explained in
Ref.~\cite{Gavela:2009cd} the observation of several radiative
decays would allow to overconstrain this model. In the type-I
seesaw models, other channels like $\mu \rightarrow eee$ or $\tau
\rightarrow 3l$ are expected to be more suppressed than the
radiative decays, see e.g.~Ref.~\cite{Antusch:2006vwa}.

\section{Softly broken $L$ case: $Y'_\alpha=0$}

At the end of the previous section, by setting $\mu_1$ to 0 in
Fig.~1, we numerically considered a simple case where the $C\!P$
asymmetry is proportional to the $Y'$ couplings. It is interesting
to discuss what happens if we take $Y'=0$ instead. This is the
situation of the usual inverse seesaw models
\cite{Wyler:1982dd}-\cite{GonzalezGarcia:1988rw} where $L$ is
assumed to be softly broken, and where in full generality $n$ pairs
of $N_{1,2}$ are usually assumed. In such a framework the Yukawa
coupling matrix $Y$ in Eqs.~(\ref{LC})-(\ref{Massmatrix}) is an
$n\times 3$ matrix, $\mu_{1,2}$ are $n \times n$ ma\-tri\-ces and
$Y'$ is a null $n\times 3$ matrix. It must be stressed that in this
case too leptogenesis can be compatible with observable flavour
violation rates. This can be seen from the $n=1$ case above which,
through the first term of Eq.~(\ref{CPasymm}), gives the $C\!P$
asymmetry \bea \eps_{1\al} =\eps_{2\al} \simeq -
{|Y_{\alpha}|^2\over 4\pi}\sin\alpha \,{\mu_1\mu_2\over 2\mu
M}\,f_{\rm self} \stackrel{\delta\ll 1}{\simeq}\,
 -\frac{|Y_\alpha|^2}{16 \pi}\, \frac{\mu_1 \mu_2}{\mu^2} \, \sin \alpha
 \,,
\label{CPasymsoft} \eea where Eq.~(\ref{selfCP}) was used in the
last equality. It is worth mentioning again that the $C\!P$
asymmetry is not suppressed by the small $L$-violating entries in
the limit of small splitting $\delta \ll 1$. The larger are the
$Y_\alpha$ couplings, the larger is the $\mu \rightarrow e \gamma$
rate and the larger is the $C\!P$ asymmetry. Note that a
non-vanishing value of both $\mu_1$ and $\mu_2$ is required in order
to have a non-zero $C\!P$ asymmetry, since otherwise the phase
$\alpha$ can be rotated away. As for the washout, it is again
suppressed as soon as $\delta\ll 1$,
Eq.~(\ref{interferencefinalN1}).

The generalization to the inverse seesaw case with $n>1$ pairs of
$N_{1,2}$ is straightforward. In this case the $N_1$'s and $N_2$'s
form $n$ pairs of quasi-degenerate states and, as in
Eqs.~(\ref{interferencefinalN1}) and (\ref{CPasymsoft}), each pair
gives an unsuppressed $C\!P$ asymmetry and a small washout as soon
as the mass splitting is smaller than the decay width. Note that
when $Y'_\alpha=0$, $n\geq 2$ is required in order to give at least
2 neutrino masses. To consider this case in full details is beyond
the scope of this paper because it involves many more parameters.

\section{Conclusion}

We have shown that, in the context of approximately $L$-conserving
seesaw models, it is rather easy to generate large lepton flavour
effects in agreement with successful leptogenesis. These models
provide all the necessary ingredients. They involve large Yukawa
couplings without leading to large neutrino masses, as required to
induce large rates. They predict a small right-handed neutrino mass
splitting
protected from large radiative corrections by the approximate
$L$-symmetry. If the splitting is sufficiently small, $\delta \ll
1$, the washout is suppressed, proportionally to the small
$L$-violating entries. And, finally, from the same condition,
$\delta \ll 1$, the suppression of the $C\!P$ asymmetry by the small
$L$-violating entries is compensated by its resonant behaviour.
In this way the experimental upper bound of any of the rare lepton processes $\mu\rightarrow e \gamma$, $\tau \rightarrow \mu \gamma$, $\tau \rightarrow e \gamma$ can be saturated in agreement with successful leptogenesis. We emphasize that this pattern can work with only 2 RH neutrinos and that moreover it does not require to assume any flavour symmetry. Only a lepton number assignment for the various RH neutrinos has to be assumed for lepton number to be approximately conserved.

To illustrate these results we have considered a minimal 2 RH neutrino model which has the additional virtue that from the knowledge of the neutrino mass matrix one can predict the rates of any rare lepton process up to an overall normalization scale. Therefore this model can be overconstrained. To our knowledge, if one does not assume any extra particles beside the RH neutrinos, there is no neutrino mass and leptogenesis seesaw model more testable than this one.

Our results also apply to the usual inverse seesaw models where $L$
is softly broken, with $n$ pairs of fermion singlets and $Y'=0$.
Similarly it also applies to the 3 RH neutrino case where, if $L$
conservation is approximate, two RH neutrinos are naturally
quasi-degenerate. In particular they apply to the 3 RH neutrino
models considered in
Refs.~\cite{Kersten:2007vk,Abada:2007ux,Gavela:2009cd}. For the
third RH neutrino, one has just to make sure that its mass is either
sizably heavier than the two others or that at least one of its
Yukawa coupling is suppressed enough.

Finally it is worth mentioning that our main result about the washout suppression also applies to the Type-III seesaw models with the same Yukawa coupling matrix and
heavy state mass matrix.
However, leptogenesis suffers in this case from the high degree of thermalization of the fermion triplets by the gauge interactions~\cite{Hambye:2003}.
It would be interesting to study more extensively this case to determine which region of the parameter space is allowed.

\section*{Acknowledgements}

We thank F.~Bonnet, B.~Gavela, P.~Hernandez and D.~Hernandez for useful discussions.
SB is partially supported by the Maryland Center for Fundamental Physics.
The work of TH and FXJM is supported by
the FNRS-FRS, the IISN and the Belgian Science Policy (IAP VI-11).

\end{document}